\documentclass[sigconf,screen]{acmart}

\renewcommand\footnotetextcopyrightpermission[1]{} 
\usepackage{booktabs}
\usepackage{microtype}

\AtBeginDocument{%
  \providecommand\BibTeX{{%
    \normalfont B\kern-0.5em{\scshape i\kern-0.25em b}\kern-0.8em\TeX}}}

\acmConference[ESEC/FSE '20]{Proceedings of the 28th ACM Joint European Software Engineering Conference and Symposium on the Foundations of Software Engineering}{November 8--13, 2020}{Virtual Event, USA}

\begin{document}
\begin{sloppy}

\title{Beyond Accuracy: Assessing Software Documentation Quality}

\author{Christoph Treude}
\email{christoph.treude@adelaide.edu.au}
\affiliation{%
  \institution{University of Adelaide}
  \city{Adelaide}
  \state{SA}
  \country{Australia}
}

\author{Justin Middleton}
\email{jamiddl2@ncsu.edu}
\affiliation{%
  \institution{North Carolina State University}
  \city{Raleigh}
  \state{NC}
  \country{United States}
}

\author{Thushari Atapattu}
\email{thushari.atapattu@adelaide.edu.au}
\affiliation{%
  \institution{University of Adelaide}
  \city{Adelaide}
  \state{SA}
  \country{Australia}
}

\begin{abstract}
Good software documentation encourages good software engineering, but the meaning of ``good'' documentation is vaguely defined in the software engineering literature. 
To clarify this ambiguity, we draw on work from the data and information quality community to propose a framework that decomposes documentation quality into ten dimensions of structure, content, and style.
To demonstrate its application, we recruited technical editors to apply the framework when evaluating examples from several genres of software documentation.
We summarise their assessments---for example, reference documentation and README files excel in quality whereas blog articles have more problems---and we describe our vision for reasoning about software documentation quality and for the expansion and potential of a unified quality framework.
\end{abstract}



\maketitle

\section{Introduction and Motivation}

High-quality software documentation is crucial for software development, comprehension, and maintenance, but the ways that documentation can suffer poor quality are numerous.
For example, Aghajani et al.~\cite{aghajani2019software} have designed a taxonomy of 162 documentation issue types, covering information content, presentation, process-related matters, and tool-related matters.
Additionally, because software documentation is written in informal natural language which is inherently ambiguous, imprecise, unstructured, and complex in syntax and semantics, its quality can often only be evaluated manually~\cite{khamis2013applying}. 
The assessment depends on context for some quality attributes (e.g., task orientation, usefulness), it mixes content and medium for others (e.g., visual effectiveness, retrievability), and sometimes it requires looking beyond the documentation itself (e.g., accuracy, completeness). 
While most related work has focused on documentation accuracy and completeness (e.g., for API documentation~\cite{zhong2013detecting, dagenais2014using}), the fact that many developers are reluctant to carefully read documentation~\cite{zhong2009inferring} suggests that documentation suffers from issues beyond accuracy.
As such, we lack a comprehensive framework and instruments for assessing software documentation quality beyond these characteristics. 

In this work we adapt quality frameworks from the data and information quality community to the software engineering domain. 
We design a survey instrument to assess software documentation quality from different sources. 
A pilot study with four technical editors and 41 documents related to the R programming language provides initial evidence for the strengths and weaknesses of different genres of documentation (blog articles, reference documentation, README files, Stack Overflow threads, tutorials) based on the ten dimensions of our software documentation quality framework.

The contributions of this work are:
\begin{itemize}
    \item A ten-dimensional framework for asking questions about software documentation quality, 
    \item A partially validated survey instrument to evaluate document quality over multiple documentation genres, and
    \item A vision for the expansion of a unified quality framework through further experimentation.
\end{itemize}

\section{Background and Related Work}

\label{sec:background}


The most related piece of work to this paper is the seminal 1995 article ``Beyond Accuracy: What Data Quality Means to Data Consumers'' by Wang and Strong~\cite{wang1996beyond}. We follow the same \textit{beyond-accuracy} approach for the domain of software documentation. 

\paragraph{Defective Software Documentation}


Defect detection tools have been widely investigated at the code level, but very few studies focus on defects at the document level~\cite{zhou2019drone}. The existing approaches in the documentation space investigate inconsistencies between code and documentation. In one of the first such attempts, Tan et al.~\cite{tan2012tcomment} presented @tcomment for testing Javadoc comments related to null values and exceptions. \textsc{DocRef} by Zhong and Su~\cite{zhong2013detecting} detects API documentation errors by seeking out mismatches between code names in natural-language documentation and code. AdDoc by Dagenais and Robillard~\cite{dagenais2014using} automatically discovers documentation patterns which are defined as coherent sets of code elements that are documented together. Also aimed at inconsistencies between code and documentation, Ratol and Robillard~\cite{ratol2017detecting} presented Fraco, a tool to detect source code comments that are fragile with respect to identifier renaming.

Wen et al.~\cite{wen2019large} presented a large-scale empirical study of code-comment inconsistencies, revealing causes such as deprecation and refactoring. Zhou et al.~\cite{zhou2018automatic, zhou2019drone} contributed a line of work on detecting defects of API documents with techniques from program comprehension and natural language processing. 
They presented DRONE to automatically detect directive defects and recommend solutions to fix them. 
And in recent work, Panthaplackel et al.~\cite{panthaplackel2020learning} proposed an approach that learns to correlate changes across two distinct language representations, to generate a sequence of edits that are applied to existing comments to reflect code modifications.

To the best of our knowledge, little of the existing work targets the assessment or improvement of the quality of software documentation beyond accuracy and completeness; one example is Pl\"{o}sch and colleagues' survey on what developers value in documentation~\cite{plosch2014value}. 
We turn to related work from the data and information quality community to further fill this gap.

\paragraph{Information Quality} 

In their seminal 1995 work ``Beyond Accuracy: What Data Quality Means to Data Consumers'', Wang and Strong~\cite{wang1996beyond} conducted a survey to generate a list of data quality attributes that capture data consumers' perspectives. These perspectives were grouped into accuracy, relevancy, representation, and accessibility, and they contained a total of 15 items such as believability, reputation, and ease of understanding. A few years later, Eppler~\cite{eppler2001generic} published a similar list of dimensions which contained 16 dimensions such as clarity, conciseness, and consistency.

These two sets of dimensions built the starting point into our investigation of dimensions from the data and information quality community which might be applicable to software documentation. Owing to the fact that software documentation tends to be disseminated over the Internet, we further included the work of Knight and Burn~\cite{knight2005developing}, who discussed the development of a framework for assessing information quality on the World Wide Web and compiled a list of 20 common dimensions for information and data quality from previous work, including accuracy, consistency, and timeliness.

\section{Software Documentation Quality}

\begin{table*}
\caption{Dimensions of software documentation quality}
\label{tab:framework}
\begin{tabular}{@{\hspace{0em}}l@{\hspace{.6em}}l@{\hspace{.6em}}l@{\hspace{0em}}}
\toprule
Dimension & Question & Source \\
\midrule 
\textsc{Quality} & How well-written is this document (e.g., spelling, grammar)? & question from~\cite{pitler2008revisiting} \\
\textsc{Appeal} & How interesting is it? & question from~\cite{pitler2008revisiting} \\
\textsc{Readability} & How easy was it to read? & `accessibility'~\cite{eppler2001generic, knight2005developing} \\
\textsc{Understandability} & How easy was it to understand? & `ease of understanding'~\cite{wang1996beyond}; question from~\cite{pitler2008revisiting} \\
\textsc{Structure} & How well-structured is the document? & `navigation' from~\cite{knight2005developing} \\
\textsc{Cohesion} & How well does the text fit together? & question from~\cite{pitler2008revisiting} \\
\textsc{Conciseness} & How succinct is the information provided? & `concise', `amount of data'~\cite{knight2005developing, wang1996beyond}; `conciseness'~\cite{eppler2001generic} \\
\textsc{Effectiveness} & Does the document make effective use of technical vocabulary? &  `vocabulary'~\cite{pitler2008revisiting} \\
\textsc{Consistency} & How consistent is the use of terminology? & `consistency'~\cite{eppler2001generic, knight2005developing} \\
\textsc{Clarity} & Does the document contain ambiguity? & `clarity'~\cite{eppler2001generic}; `understandability'~\cite{knight2005developing} \\
\bottomrule
\end{tabular}
\end{table*}

Inspired by the dimensions of information and data quality identified in related work, we designed the software documentation quality framework summarised in Table~\ref{tab:framework}. The first and last author of this paper collaboratively went through the dimensions from related work cited in Section~\ref{sec:background} to select those that (i) could apply to software documentation, (ii) do not depend on the logistics of accessing the documentation (e.g., security), and (iii) can be checked independently of other artefacts (i.e., not accuracy or completeness). As a result, we omitted dimensions such as relevancy and timeliness from Wang and Strong's work~\cite{wang1996beyond} since they require context beyond the documentation itself (e.g., relevancy cannot be assessed without a specific task in mind); and we omitted dimensions such as interactivity and speed from Eppler's work~\cite{eppler2001generic} which are characteristics of the medium rather than the content. Table~\ref{tab:framework} indicates the sources from related work for each dimension.

In addition, we formulated questions that an assessor can answer about a piece of software documentation to indicate its quality in each dimension. Our pilot study in Section~\ref{study} provides initial evidence that technical editors are able to answer these questions and shows interesting trends across genres of documentation. The questions were inspired by work on readability by Pitler and Nenkova~\cite{pitler2008revisiting} who collected readability ratings using four questions: How well-written is this article? How well does the text fit together? How easy was it to understand? How interesting is this article? Our framework contains the same questions, and complements them with one question each for the additional six dimensions. 

\section{Pilot Study} \label{study}

To gather preliminary evidence on whether experts would be able to use the proposed framework for assessing software documentation, we conducted a pilot study with technical editors and different genres in software documentation. 


\paragraph{Recruitment}

To recruit technical editors, we posted an advertisement on Upwork,\footnote{\url{https://www.upwork.com/}} a website for freelance recruitment and employment.
Our posting explained our goal of evaluating software documentation on ten dimensions and identifying weaknesses in their design.
Therefore, we required applicants to be qualified as technical editors with programming experience, and accordingly, we would compensate them hourly according to their established rates.
From this, we recruited the first four qualified applicants with experience in technical writing and development.
We refer to our participants henceforth as E1 through E4.

\paragraph{Data Collection} \label{datacollect}

Because developers write software instructions in a variety of contexts, we use a broad definition of documentation, accepting both official, formal documentation and implicit documentation that emerges from online discussion. 
We focus on the genres of software documentation that have been the subject of investigation in previous work on software documentation: reference documentation~\cite{fucci2019using}, README files~\cite{prana2019categorizing}, tutorials~\cite{petrosyan2015discovering}, blogs~\cite{parnin2013blogging}, and Stack Overflow threads~\cite{barua2014developers}. 
To better control for variation between language, we focused exclusively on resources documenting the R programming language or projects built with it.
All documentation was then randomly sampled from the following sources: 

\begin{itemize}
    \item \textbf{Reference Documentation (RD)}: from the 79 subsections of the R language manual.\footnote{\url{https://cran.r-project.org/doc/manuals/r-release/R-intro.html}}
    \item \textbf{README files (R)}: from the first 159 items in a list\footnote{\url{https://github.com/qinwf/awesome-R}, the remaining items were not software projects} of open-source R projects curated by GitHub users. 
    \item \textbf{Tutorials (T)}: from the 46 R tutorials on tutorialspoint.com.\footnote{\url{https://www.tutorialspoint.com/r/index.htm}}
    \item \textbf{Articles (A)}: from 1,208 articles posted on R-bloggers.com,\footnote{\url{https://www.r-bloggers.com/}} a blog aggregation website, within the last six months.
    \item \textbf{Stack Overflow threads (SO)}: from questions tagged ``R''.
\end{itemize}

We asked the editors to read the documents, suggest edits, and answer the ten questions listed in Table~\ref{tab:framework} using a scale from 1 (low) to 10 (high). 
From a methodological standpoint, we asked for edits in addition to the ten assessments in order to encourage and evidence reflection on the specific ways the documentation failed, beyond the general impressions of effectiveness.
These edits were collected through Microsoft Word's Review interface.
We assigned the quantity and selection of documentation according to the amount of time editors had available while trying to balance distribution across genres.
Furthermore, we tried to give each individual document to at least two editors.
We show the distribution in Table~\ref{tab:participant_assignments}.
\begin{table}[t]
    \centering
    \caption{Distribution of documentation genres to editors}
    \begin{tabular}{lrrrrrr} \toprule
        Editor & RD & R & T & A & SO & Sum  \\ \midrule
        E1 & 7 & 7 & 7 & 8 & 7 & 36 \\ 
        E2 & 6 & 6 & 5 & 6 & 6 & 29 \\ 
        E3 & 1 & 1 & 1 & 1 & 1 & 5 \\ 
        E4 & 1 & 1 & 1 & 1 & 1 & 5 \\ 
        \midrule
        Sum & 15 & 15 & 14 & 16 & 15 & 75 \\
        Union & 8 & 8 & 8 & 9 & 8 & 41 \\ \bottomrule
    \end{tabular}
    \label{tab:participant_assignments}
\end{table}
Overall, the editors worked on 41 documents.
Seven documents were assessed by only one editor, but the rest were by two; therefore, we had 75 records of document evaluations.


\paragraph{Results}

We display the average dimension rating per genre of documentation in Table~\ref{tab:results}; for each row, the highest value is in bold, and the lowest is italicised.
For example, this table shows that the quality of the reference documentation, or how well-written it is according to the definition in Table~\ref{tab:framework}, is rated 8.1 out of 10, on average higher than all the other genres.
Meanwhile, the average blog article was rated 6.1 out of 10, the lowest.
Within dimensions, the highest ratings are distributed primarily among reference documentation and README files; tutorials received the highest in cohesion and conciseness, but only by a difference of at most 0.3 of a rating point with the two aforementioned genres.
Nevertheless, blog articles received the global lowest rating across every dimension, especially so in readability and structure.

\begin{table}[t]
\caption{Results of document rating by technical editors}
\begin{tabular}{lrrrrr} \toprule
Dimension & RD & R & T & A & SO \\ \midrule
\textsc{Quality} & \textbf{8.1} & 7.6 & 7.5 & \textit{6.1} & 6.7 \\
\textsc{Appeal} & 5.7 & \textbf{6.5} & 6.4 & \textit{5.3} & 5.8 \\
\textsc{Readability} & \textbf{6.6} & 6.4 & 6.5 & \textit{4.6} & 6.1 \\
\textsc{Understandability} & 6.9 & \textbf{7.0} & 6.9 & \textit{5.3} & 6.1 \\
\textsc{Structure} & 6.3 & \textbf{7.6} & 7.1 & \textit{3.8} & 5.8 \\
\textsc{Cohesion} & 6.7 & 6.7 & \textbf{7.0} & \textit{4.9} & 6.1 \\
\textsc{Conciseness} & 6.7 & 6.7 & \textbf{6.8} & \textit{5.3} & 6.3 \\
\textsc{Effectiveness} & \textbf{8.3} & 7.9 & 8.1 & \textit{7.1} & 7.5 \\
\textsc{Consistency} & \textbf{9.3} & 8.8 & 9.0 & \textit{8.5} & 8.6 \\
\textsc{Clarity} & 8.6 & \textbf{8.7} & 7.6 & \textit{6.3} & 7.7 \\ \bottomrule
\end{tabular}
\label{tab:results}
\end{table}


Different dimensions do not seem to have equal distributions.
For example, reference documentation has a 9.3 in consistency, but the lowest score is an 8.5 for blog articles.
These are nevertheless high scores on average.
Appeal also has a small spread, from 6.5 in README files to 5.3 in blog articles, but a lower average overall across dimensions.
Structure, on the other hand, varies from 7 (README files) to 3.8 (blog articles), demonstrating a larger interaction with genres.

\paragraph{Reflection}
Our findings suggest a trend in rating samples from certain genres highly while disapproving others.
The relatively high rating of reference documentation and README files may be explained by their typical origins within the software product and team, 
whereas blog articles and Stack Overflow posts can be published more freely~\cite{parnin2012crowd}.
Nevertheless, the result that blog articles received the lowest in every category did surprise us, especially as we designed the framework to reduce the impact of context.
There may be many factors involved in these results.
For one, the corpora from which we randomly sampled may have different underlying distributions on our dimensions.
Our reference documentation was sampled from larger related products, whereas manual inspection of the randomly sampled blog articles did evidence a broad variety of R-related topics, such as discussions of R package updates or write-ups and reflections on personal projects.
Furthermore, as with any participant study, the soundness of our results depends on the accuracy with which we communicated our ideas and the participants understood and enacted them.
Editors may approach different genres with different preconceptions; an 8 rating for a README file may not be the same as an 8 rating for a Stack Overflow thread.

\paragraph{Technical Challenges} As noted previously, we asked editors to make edits on document copies in Microsoft Word as well, hoping to gain insight into technical editing strategies for software documentation.
We obtained over 4,000 edit events across all 75 documents, ranging from reformattings and single character insertions to deep paragraph revisions.
However, we encountered several obstacles when attempting to analyse this data.
First, when copying online documents to Microsoft Word, interactions and images in hypertext documents were distractingly reformatted for a linear document.
As a result, several of the edits addressed changes to structure that did not exist when viewing the document in the browser instead.
Furthermore, many edits were not recorded as intended---for example, replacing words with synonyms that do not make sense in a programming context.
Although we can recreate changes to the document text through before-and-after comparison, small edits blend together and large edits confound the differencing algorithm.
Because the editor's original intent is lost, any coding scheme applied over it becomes less secure.
Due to these circumstances, we decided against drawing conclusions from this data.

\section{Impact and Future Work}

Accuracy of software documentation is a necessary but not sufficient condition for its success. To empower software developers to make effective use of software documentation, it must be carefully designed in appearance, content, and more. Our vision, then, is to provide a more precise definition to better explore and transform what it means for software documentation to be of high quality, along with a research agenda enabled by such a quality framework. To that end, we have drawn from seminal work in the information and data quality communities to design a framework consisting of ten dimensions for software documentation quality.

Furthermore, our research agenda proposes future work to evaluate the impact of each dimension with end-users and improve the framework.
For one, we can verify the assessments of technical editors by introducing end-users to different versions of the documents (e.g., edited by technical editors based on quality criteria and original version) and observing their use.
Another direction is to explore trade-offs between quality attributes, such as whether readability outweighs structure in terms of document usability, and to further disambiguate similar dimensions (e.g., structure and cohesion). 
We will further revisit the edits from the technical editors to extract surface-level (e.g., word count) and deep-level (e.g., cohesion of adjacent sentences) lexical, syntactic, and discourse features to build classification models for predicting documentation quality. 
Such classification models can be used to assess and rank the quality of software documentation.

We believe these efforts are important because of the volume of software documentation on the web. 
A simple Google search related to software development will return documentation that matches the query without explicitly considering the quality of the material. 
For example, a query on `reading CSV file in r' returns blog articles and tutorials as the top results, yet our preliminary results demonstrate that on average, blog articles are ranked worst across all ten quality dimensions.
This is not to say that blog articles are essentially faulty and should be abandoned moving forward; rather, we hope to spur reflection on how end-users interact with blog articles and how each of our dimensions manifest uniquely under the genre's constraints, while nevertheless using what we currently know to emphasise more useful results. 
Therefore, applying the framework can influence guidelines and recommendation systems for documentation improvement as well as automatic assessing and ranking systems for navigating the large volumes of documentation and emphasising high-quality documentation. 

\section*{Acknowledgements}

The authors thank Emerson Murphy-Hill and Kathryn T.~Stolee for their contributions to this work. This research was undertaken, in part, thanks to funding from a University of Adelaide -- NC State Starter Grant and the Australian Research Council's Discovery Early Career Researcher Award (DECRA) funding scheme (DE180100153). This work was inspired by the International Workshop series on Dynamic Software Documentation, held at McGill's Bellairs Research Institute.


\end{sloppy}
\end{document}